\documentclass{article}
\usepackage{graphicx}
\usepackage{amsmath}
\usepackage[ruled]{algorithm2e}
\usepackage{fullpage}
\usepackage{authblk}

\SetAlFnt{\small}
\SetAlCapFnt{\small}
\SetAlCapNameFnt{\small}
\SetAlCapHSkip{0pt}
\IncMargin{-\parindent}
\SetKwProg{Fn}{Function}{}{end}
\SetKwFunction{HILBERTENCODE}{encodeHilbert3D}
\SetKwFunction{HILBERTDECODE}{decodeHilbert3D}

\title{Algorithms for Encoding and Decoding 3D Hilbert Orderings}
\author{David Walker}

\affil{UTC Research Institute, University of Tennessee at Chattanooga, TN 37403}
\date{August 2023}

\begin{document}

\maketitle

\section{Introduction}
\begin{figure}[ht]
  \centering
  \includegraphics[width=0.5\linewidth]{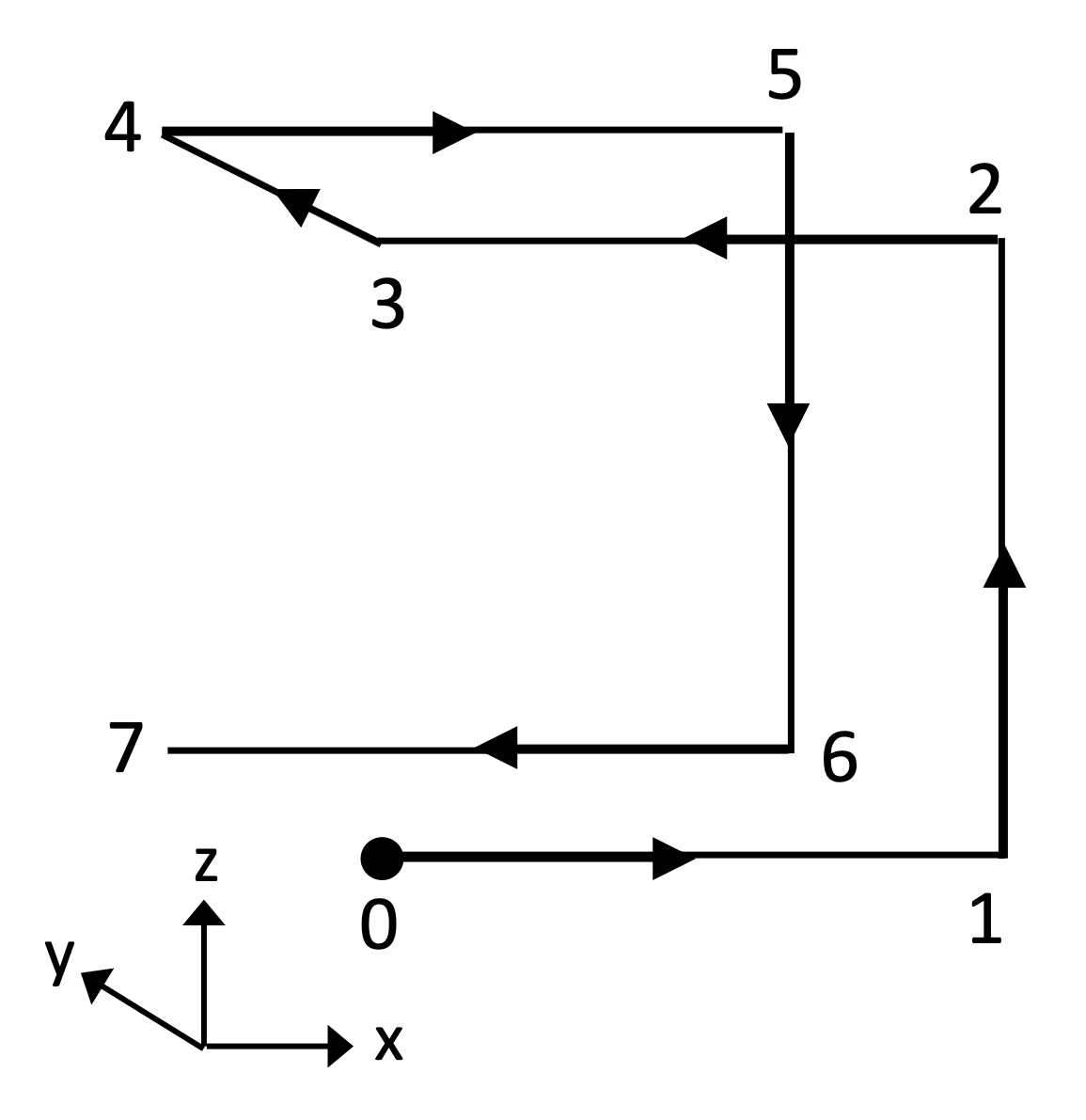}
  \caption{The base ordering for a 3D Hilbert curve. Each of the 8 locations are labelled by the Hilbert index and neighbors are connected by a path of unit length. The start is indicated by a filled circle at $(0,0,0)$, and the path direction is shown by arrows.}
  \label{fig:base}
\end{figure}

This paper presents algorithms and pseudocode for encoding and decoding 3D Hilbert orderings, which have application in a number of areas because of their locality properties. This problem has been previously considered, for example by~\cite{Haverkort2017,LiuSchrak1997}, but to the best of our knowledge this is the first clear computational presentation in terms of algorithms and pseudocode. The general approach taken is similar to that of \cite{Chen2006} which deals with the 2D case.
A 3D Hilbert ordering of depth $r$ passes once through each location $(x,y,z)$ in a $M\times M\times M$ data cube, where $M=2^r$ and $x$, $y$, and $z$ are integer indices from 0 to $M-1$. Successive locations in a Hilbert ordering differ by addition or subtraction of one of the basis vectors $(1,0,0)$, $(0,1,0)$, and $(0,0,1)$. For $r=1$ the ordering is shown in Fig.~\ref{fig:base}, and is referred to as the \emph{base ordering}.

\section{Generating a Hilbert Curve Incrementally}

In general, a Hilbert curve can be represented by a Lindenmayer system (or \emph{L-system)} in terms of parallel rewrite rules \cite{Lindenmayer1968,Prusinkiewicz1996}. 
A 3D Hilbert curve can be represented as an L-system with the following rewrite rule:
\begin{eqnarray*}
X&\rightarrow& \wedge<XF\;\wedge<XFX-F\;\wedge>>XFX\vee F\;
+>>XFX-F>X->
\label{eqn:hilbert3d}
\end{eqnarray*}
where the meanings of the symbols are given in Table~\ref{tab:symbols}. Applying this rewrite rule  recursively $r$ times draws a Hilbert curve of depth $r$. The rotations in Table~\ref{tab:symbols} are interpreted relative to the current heading, and two other mutually orthogonal axes,  encoded in an orientation matrix. The algorithm proceeds by recursively processing the rewrite rule from left to right and updating the orientation matrix each time a symbol from Table~\ref{tab:symbols} is encountered by post-multiplying it by the corresponding rotation matrix. When the symbol F is encountered one of the basis vectors is added or subtracted to the current $(x,y,z)$ location, depending on the current orientation of the axes~\cite{AlKharusiWalker2019}. This generates the next location in the Hilbert path through the data cube and thus follows the Hilbert curve, starting at index 0 and location $(0,0,0)$.
\begin{table}[hbt]
\centering
\begin{tabular}{|l|c|c|c|c|c|c|}
\hline
Symbol & $+$ & $-$ & $\wedge$ & $\vee$ & $<$ & $>$\\
Meaning  & Yaw 90$^\circ$ & Yaw -90$^\circ$ & Pitch 90$^\circ$ & Pitch -90$^\circ$ & Roll 90$^\circ$ & Roll -90$^\circ$\\
\hline
\end{tabular}
\caption{Meaning of the symbols in the rewrite rule for a 3D Hilbert curve.}
\label{tab:symbols}
\end{table}

\section{Mapping Between a  Hilbert Index and a Data Location}
\begin{figure}[ht]
  \centering
  \includegraphics[width=0.8\linewidth]{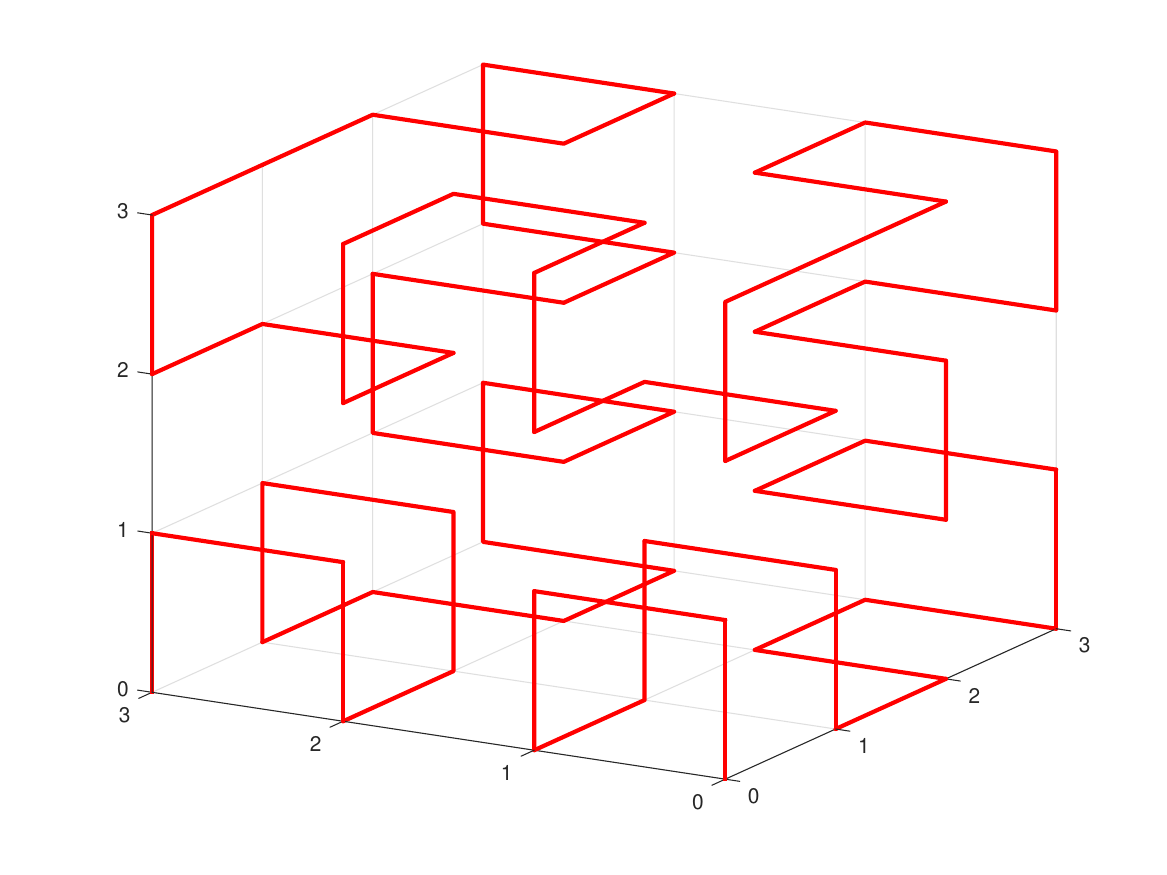}
  \caption{A 3D Hilbert ordering of depth 2. The Hilbert path starts at $(0,0,0)$ and ends at $(0,3,0)$, corresponding to Hilbert indices 0 and 63, respectively.}
  \label{fig:hilbert3d}
\end{figure}

\begin{figure}[ht]
  \centering
  \vspace*{-0.5in}
  \includegraphics[width=0.8\linewidth, trim= {0 4cm 0 4cm}, clip]{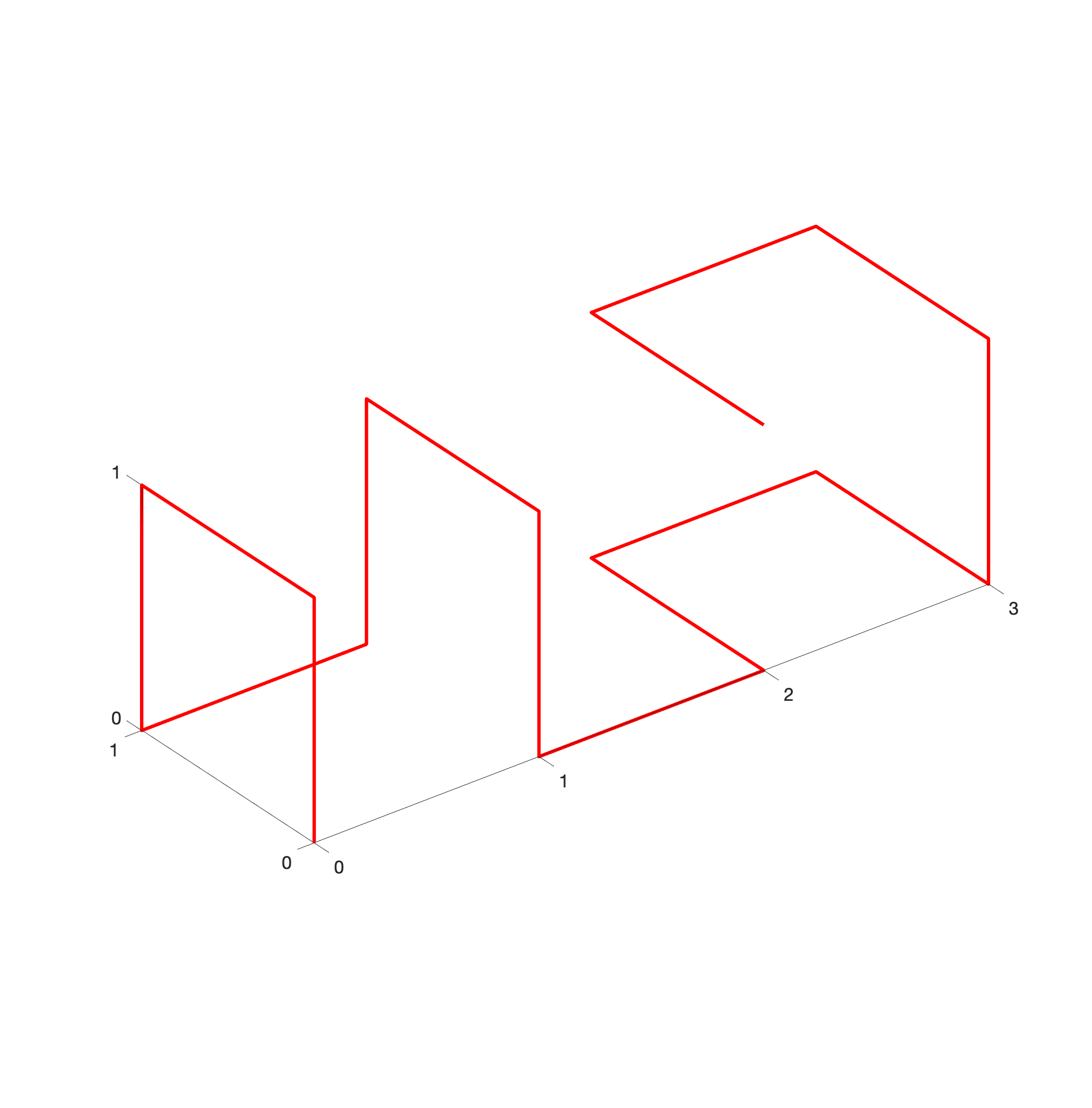}
  \vspace*{-0.6in}  
  \caption{Octants 0 and 1 of a 3D Hilbert ordering of depth 2.}
  \label{fig:octants0and1}
\end{figure}

There are many applications in which it is necessary to find the location of an arbitrary Hilbert index (or {\it vice versa}) --  for example, in stencil-based computations. 
Given a Hilbert ordering of depth $r$, an ordering of depth $r+1$ is obtained by dividing the data cube into octants, which are labelled as in the base ordering. The Hilbert ordering of depth $r$ is replicated in each octant and rotated about the coordinate axes. The rotation is such that the last location in octant $P$ is connected to the first location in octant $P+1$ by addition or subtraction of one of the basis vectors $(1,0,0)$, $(0,1,0$, and $(0,0,1)$, for $P=0,1,\ldots,6$. Figure~\ref{fig:hilbert3d} shows a 3D Hilbert curve of depth 2. The first 8 locations, starting at $(0,0,0)$, occupy octant 0, the next 8 locations occupy octant 1, and so on. For clarity, Fig.~\ref{fig:octants0and1} shows the first 16 locations making up octants 0 and 1. The first octant contains the base ordering rotated clockwise by $\pi/2$ radians about the y axis, and by $\pi/2$ radians clockwise about the z axis. Similarly, octant 1 contains the base ordering rotated by $\pi/2$ radians about the x axis and then by the same amount about the z axis, where positive rotation angles are counterclockwise. The last location in octant 0 is connected to the first location in octant 1 by the vector $(1,0,0)$. The replicated base orderings in the other octants are rotated and connected in similar ways.

The rotations that transform the base ordering into the orderings within each octant can be expressed in terms of counterclockwise rotations about the x, y, and z axes by $\pi/2$ radians, represented by the matrices $X$, $Y$, and $Z$, where:
\[
X=\begin{pmatrix}1 &0& 0 \\0& 0& -1\\0& 1& 0\end{pmatrix},\qquad
Y=\begin{pmatrix}0 &0& 1 \\0& 1& 0\\-1& 0& 0\end{pmatrix},\qquad
Z=\begin{pmatrix}0 &-1& 0 \\1& 0& 0\\0& 0& 1\end{pmatrix}.
\]
The transformation matrix for each octant is given by:

\begin{table}[ht]
\centering
\begin{tabular}{|c|cccccccc|}
\hline &&&&&&&& \\[-2.5ex]
Octant&0&1&2&3&4&5&6&7\\
\hline &&&&&&&& \\[-2.2ex]
Rotations&$Z^TY^T$&$ZX$&$ZX$&$Y^2$&$Y^2$&$Z^TX^T$&$Z^TX^T$&$XZ$\\
\hline
\end{tabular}
\caption{Transformation matrices for each octant.}
\label{tab:matrices}
\end{table}

\subsection{Encoding Algorithm}
Given a location $(x,y,z)$ in the data cube, the encoding algorithm determines the corresponding Hilbert path index, $h$. The encoding algorithm evaluates each octal digit of $h$ in turn, starting with the most significant. In general, $h$ consists of $r$ octal digits for a Hilbert path of depth $r$, and at this resolution the number of data points in each direction is $2w$, where $w=2^{r-1}$. The encoding algorithm determines the octant containing $(x,y,z)$ at this resolution by evaluating $x/w$, $y/w$, and $z/w$, with the octant being indexed 0 to 7 as in the base ordering in Fig.~\ref{fig:base}. This octant index is the most significant octal digit of $h$. The encoding algorithm next applies the following series of transformations to $(x,y,z)$:
\begin{enumerate}
    \item Move the origin of the coordinate system to position $(0,0,0)$ of the octant by: subtracting $w$ from $x$ if the octant index is 1, 2, 5, or 6; subtracting $w$ from $y$ if the octant index is 4, 5, 6,  or 7; and, subtracting $w$ from $z$ if the octant index is 2, 3, 4, or 5.
    \item Move the origin of the coordinate system to the center of the octant by subtracting $w_*$ from $x$, $y$, and $z$, where $w_*=(w-1)/2$.
    \item Transform the resulting $(x,y,z)$ vector using the inverse of the corresponding matrix in Table~\ref{tab:matrices}.
    \item Move the origin of the coordinate system back to position $(0,0,0)$ in the octant by adding $w_*$ to $x$, $y$, and $z$.
\end{enumerate}
Setting $w=w/2$ and repeating the above steps will evaluate the next octal digit of $h$. The process is repeated a total of $r$ times to evaluate all $r$ octal digits of $h$. The transformations encapsulating the above steps are shown in Table~\ref{tab:encoding}.

\begin{table}[ht]
    \centering
    \begin{tabular}{cccc}
    \hline
    Octant&$x_{new}$&$y_{new}$&$z_{new}$\\
    \hline
    0 & $z$ & $x$ & $y$\\
    1 & $y$ & $z$ & $x-w$\\  
    2 & $y$ & $z-w$ & $x-w$\\
    3 & $w-x-1$ & $y$ & $2w-z-1$\\ 
    4 & $w-x-1$ & $y-w$ & $2w-z-1$\\
    5 & $2w-y-1$ & $2w-z-1$ & $x-w$\\ 
    6 & $2w-y-1$ & $w-z-1$ & $x-w$\\
    7 & $z$ & $w-x-1$ & $2w-y-1$\\    
    \hline 
    \end{tabular}
    \caption{Hilbert encoding update rules.}
    \label{tab:encoding}
\end{table}

An optimization can be made when some of the most significant octal digits of $h$ are zero, since the corresponding octant for such digits is known to be 0. Denote by $r_{min}$ the number of octal digits of $h$, ignoring any leading zeroes. Thus, $r_{min}=\log_2{(\max(x,y,z))} + 1$ (or 1 if $(x,y,z)=(0,0,0)$). The transformation for octant 0 in Table~\ref{tab:encoding} is applied $r-r_{min}$ times to $(x,y,z)$, and then the steps enumerated above are applied $r_{min}$ times to evaluate the remaining $r_{min}$ digits of $h$, starting with $w=2^{r_{min}-1}$. Noting that applying the transformation for octant 0 three times gives the identity transformation, then if $t=(r-r_{min})\pmod 3$, the transformation can be applied $r-r_{min}$ times by applying it just $t$ times. Thus, if $t=1$ then $(x,y,z)\rightarrow(z,x,y)$ and if $t=2$ then $(x,y,z)\rightarrow(y,z,x)$. Algorithm~\ref{alg:encoding} shows the encoding algorithm with this optimization.

\begin{algorithm}[ht]
\SetAlgoNoLine
\DontPrintSemicolon
\Fn{\HILBERTENCODE{$x,y,z,r$}}{
\KwIn{Location $(x,y,z)$ and Hilbert depth, $r$.}
\KwOut{Hilbert index, $h$.}
Evaluate $r_{min}$\;
Evaluate $t=(r-r_{min})\pmod{3}$\;
\uIf{$t=1$}{
    $(x,y,z)=(z,x,y)$\;}
\ElseIf{$t=2$}{
    $(x,y,z)=(y,z,x)$\;}
Initiate $h=0$\;
Initiate $w=2^{r_{min}-1}$\;
\For{$k=r_{min}$ \KwTo{$1$}}
{Use $w$ to evaluate octant, $o$\;
$h=8*h+o$\;
Apply encode update rule for octant $o$ to $(x,y,z)$\;
$w=w/2$\;
}
return $h$\;
}
\caption{Algorithm for encoding a location $(x,y,z)$ as a Hilbert index.}
\label{alg:encoding}
\end{algorithm}

As an example, consider location $(3,3,1)$ for a Hilbert ordering of depth $r=2$. Setting $w=2^{r-1}=2$, this location is in octant 6, which is therefore the leading octal digit of the Hilbert index, $h$. Applying the update rule for octant 6 from Table~\ref{tab:encoding}, $(x,y,z)$ becomes $(0,0,1)$ and $w=1$. This lies in octant 3, which is the next most significant octal digit of $h$. This completes the encoding so $h=63$ in octal, or 51 in decimal.

\subsection{Decoding Algorithm}

Given a Hilbert path index, $h$ and a depth $r$, the decoding algorithm determines the corresponding location $(x,y,z)$ in the data cube.  The decoding algorithm  executes the steps of the encoding algorithm in reverse order by examining each octal digit of $h$ in turn, starting with the least significant. The algorithm first finds the least significant octal digit of $h$ which gives the index of the octant of the data cube containing the corresponding location. The location $(x,y,z)$ is then initialized according to the base ordering in Fig.~\ref{fig:base}. The least significant octal digit of $h$ is then discarded by replacing $h$ with $h/8$. The value of $w$ is set to 2 and the algorithm proceeds as follows:
\begin{enumerate}
    \item Let $o$ be the least significant octal digital of the current $h$. 
    \item Move the origin of the coordinate system to the center of  octant $o$ by subtracting $w_*$ from $x$, $y$, and $z$, where $w_*=(w-1)/2$. 
    \item Transform the resulting $(x,y,z)$ vector using the corresponding matrix in Table~\ref{tab:matrices}.    
    \item Move the origin of the coordinate system back to position $(0,0,0)$ in the octant by adding $w_*$ to $x$, $y$, and $z$.    
    \item Move the origin of the coordinate system to position $(0,0,0)$ of  octant 0 by: adding $w$ to $x$ if  $o$ is 1, 2, 5, or 6; adding $w$ to $y$ if $o$ is 4, 5, 6,  or 7; and, adding $w$ to $z$ if $o$ is 2, 3, 4, or 5.
\end{enumerate}
The values $w$ and $h$ are updated to $2w$ and $h/8$, respectively, and the above steps are repeated until $h=0$. Some of the leading higher order octal digits of the original value of $h$ may be zero and it is necessary to perform the corresponding transformations for octant 0 by further updating $(x,y,z)$. Denoting the number of octal digits in $h$ by $r_{min}$ (ignoring any leading zeroes) and letting $t=(r-r_{min})\pmod{3}$, the last step of the algorithm updates $(x,y,z)$ to $(y,z,x)$ if $t=1$, and to $(z,x,y)$ if $t=2$. The transformations encapsulating the above steps are shown in Table~\ref{tab:decoding}. Algorithm~\ref{alg:decoding} shows the decoding algorithm.

\begin{table}[ht]
    \centering
    \begin{tabular}{cccc}
    \hline
    Octant&$x_{new}$&$y_{new}$&$z_{new}$\\
    \hline
    0 & $y$ & $z$ & $x$\\
    1 & $z+w$ & $x$ & $y$\\  
    2 & $z+w$ & $x$ & $y+w$\\
    3 & $w-x-1$ & $y$ & $2w-z-1$\\ 
    4 & $w-x-1$ & $y+w$ & $2w-z-1$\\
    5 & $z+w$ & $2w-x-1$ & $2w-y-1$\\ 
    6 & $z+w$ & $2w-x-1$ & $w-y-1$\\
    7 & $w-y-1$ & $2w-z-1$ & $x$\\    
    \hline 
    \end{tabular}
    \caption{Hilbert decoding update rules.}
    \label{tab:decoding}
\end{table}

\begin{algorithm}[ht]
\SetAlgoNoLine
\DontPrintSemicolon
\Fn{\HILBERTDECODE{$x,y,z,r$}}{
\KwIn{Hilbert index, $h$, and depth, $r$}
\KwOut{Location $(x,y,z)$.}
Evaluate $o$, the least significant octal digit of $h$\;
Use $o$ to initiate location $(x,y,z)$\;
Initiate $w=2$\;
$h=h/8$\;
\While{$h$}
{Extract $o$, the least significant octal digit from $h$\;
Apply decode update rule for octant $o$ to $(x,y,z)$\;
$h=h/8$\;
$w=2w$\;
}
Evaluate $r_{min}$\;
Evaluate $t=(r-r_{min})\pmod{3}$\;
\uIf{$t=1$}{
    $(x,y,z)=(y,z,x)$\;}
\ElseIf{$t=2$}{ 
    $(x,y,z)=(z,x,y)$\;}
return $(x,y,z)$\;
}
\caption{Algorithm for decoding a Hilbert index $h$ as a location $(x,y,z)$.}
\label{alg:decoding}
\end{algorithm}

As an example, consider Hilbert index $h=37$ for an ordering of depth 2. The least significant octal digit of $h$ is 5, so $(x,y,z)$ is initialized to $(1,1,1)$. Next initialize $w$ to 2 and set $h=h/8=4$ of which the least significant octal digit is $o=4$. Applying the decode update rule for octant 4 gives $(x,y,z)=(0,3,2)$. Dividing $h$ by 8 gives $h=0$ so the {\tt while} loop terminates. Thus, since $t=0$, the algorithm is finished, and Hilbert index 37 is at location $(0,3,2)$.

\section{Conclusion}
This paper has presented algorithms for encoding and decoding 3D Hilbert orderings. Future work will compare the performance of this approach with the technique of precomputing and storing the mapping between location and Hilbert index in arrays for subsequent use in an application.

\section*{Acknowledgments}
Funding in part is acknowledged from  NSF Grants 1918987, 2151020,
and 2201497, and the U.S. Department of Energy's National Nuclear
Security Administration (NNSA) under the Predictive Science Academic
Alliance Program (PSAAP-III), Award DE-NA0003966. Any opinions,
findings, and conclusions or recommendations expressed in this
material are those of the authors and do not necessarily reflect
the views of the National Science Foundation, or the U.S.\hbox{}
Department of Energy's National Nuclear Security Administration.

\bibliographystyle{plain}
\bibliography{main}

\end{document}